\documentclass[10pt,preprintnumbers,amsmath,amssymb]{revtex4}
\usepackage{units}
\usepackage{bbold,eulervm}
\usepackage{graphicx}
\usepackage{dcolumn}
\usepackage{bm,bbm,mathtools,esvect}
\usepackage{slashed}

\begin{document}

\title{\LARGE\sf Quaternionic quantum harmonic oscillator\\ \vspace{5mm}}

\author{\bf SERGIO GIARDINO} 
\email{sergio.giardino@ufrgs.br}
\affiliation{\vspace{3mm} Departamento de Matem\'atica Pura e Aplicada, Universidade Federal do Rio Grande do Sul (UFRGS)\\
Avenida Bento Gon\c calves 9500, Caixa Postal 15080, 91501-970  Porto Alegre, RS, Brazil}

\begin{abstract}
\noindent 
{\bf Abstract:} In this article we obtained the harmonic oscillator solution for quaternionic quantum mechanics
($\mathbbm{H}$QM) in the real Hilbert space, both in the analytic method and in the algebraic method. The quaternionic solutions have many additional possibilities if compared to complex quantum mechanics ($\mathbbm{C}$QM), and thus there are many possible applications to these results in future research.
\end{abstract}

\maketitle
\tableofcontents

\section{\;\sf Introduction\label{I}}
Quaternions ($\mathbbm{H}$) are generalized complex numbers comprising three 
anti-commutative imaginary units, namely $i,\,j$ and $k$. If  $q\in\mathbbm{H}$, then
\begin{equation}\label{i1}
 q=x_0 + x_1 i + x_2 j + x_3 k, \qquad\mbox{where}\qquad x_0,\,x_1,\,x_2,\,x_3\in\mathbbm{R},\qquad i^2=j^2=k^2=-1.
\end{equation}
Mathematical and physical introductions to quaternions are provided elsewhere \cite{Morais:2014rqc,Rocha:2013qtt,Garling:2011zz,Dixon:1994oqc,Ward:1997qcn}, 
and we notice only that the anti-commutativity of the imaginary units makes quaternionic numbers non-commutative hyper-complexes. By way of example $ij=k=-ji$. Adopting the symplectic notation for quaternions, (\ref{i1}) becomes
\begin{equation}\label{i2}
q=z_0+z_1j,\qquad\mbox{where}\qquad z_0=x_0+x_1i\qquad\textrm{and}\qquad z_1=x_2+x_3i.
\end{equation}
In quaternion quantum mechanics ($\mathbbm{H}$QM) the quantum states are evaluated over the quanternionic numbers. Thus, quaternionic wave functions replace the usual complex wave functions in Schr\"odinger equation, and therefore the $\mathbbm{H}$QM generalizes the usual
complex quantum mechanics ($\mathbbm{C}$QM). The introduction of quaternions in quantum mechanics is not new, and  Stephen Adler's book \cite{Adler:1995qqm} contains a large extent of their development, subsumming the anti-hermitian version of $\mathbbm{H}$QM, where
 anti-hermitian Hamiltonian operators are imposed on Schr\"odinger equation. Anti-hermitian $\mathbbm{H}$QM comprises several shortcomings, 
such as the ill-defined classical limit \cite{Adler:1995qqm}. Furthermore, anti-hermitian solutions of $\mathbbm{H}$QM are few, involved, and difficult to understand physically
\cite{Davies:1989zza,Davies:1992oqq,Ducati:2001qo,Nishi:2002qd,DeLeo:2005bs,Madureira:2006qps,Ducati:2007wp,Davies:1990pm,DeLeo:2013xfa,DeLeo:2015hza,Giardino:2015iia,Sobhani:2016qdp,Procopio:2017vwa,Sobhani:2017yee,Hassanabadi:2017wrt,Hassanabadi:2017jiz,Bolokhov:2017ndw,
Cahay:2019bqp,DeLeo:2019bcw}.
We additionally point out that several applications of quaternions in quantum mechanics are not $\mathbbm{H}$QM because the anti-hermitian framework is not considered \cite{Arbab:2010kr,Brody:2011mg,Morais:2014jpm,Kober:2015bkv,Tabeu:2019cqw,Chanyal:2019gdi,Cahay:2019bqp} and the quaternions are simply an alternative way to describe specific results of $\mathbbm{C}$QM.

More recently, a novel approach  eliminated the anti-hermiticity requirement for the Hamiltonian operator in $\mathbbm{H}$QM \cite{Giardino:2018lem,Giardino:2018rhs}. Using this 
framework, several results have been obtained, including the explicit solutions of the Aharonov-Bohm effect \cite{Giardino:2016xap},
the free particle \cite{Giardino:2017yke,Giardino:2017nqs}, the square well \cite{Giardino:2020cee}, the Lorentz force \cite{Giardino:2019xwm,Giardino:2020uab} and the quantum scattering \cite{Giardino:2020ztf,Hasan:2020ekd}. Further conceptual results are the well defined classical limit \cite{Giardino:2018lem}, the virial theorem \cite{Giardino:2019xwm}, the Ehrenfest theorem  and the real Hilbert space  \cite{Giardino:2018lem,Giardino:2018rhs}. In the real Hilbert space approach, an arbitrary quaternionic wave function $\,\Psi\,$ is written in terms of the linear expansion
\begin{equation}
\Psi=\sum_{\ell=-\infty}^\infty c_\ell \Lambda_\ell,
\end{equation}
where $\,c_\ell\,$ are real coefficients and $\,\Lambda_\ell\,$ are quaternionic basis elements. We recall that in $\mathbbm C$QM the coefficients and the basis elements are both complex, and that in the anti-hermitian $\mathbbm H$QM the coefficients and the basis elements are both quaternionic. A real Hilbert space is endowed with a real valued inner product, and from \cite{Harvey:1990sca} a consistent real inner product between the quaternions  $\Phi$ and $\Psi$ is simply
\begin{equation}\label{u003}
\langle\Phi,\,\Psi\rangle= \frac{1}{2}\int dx^3\Big[\Phi\overline\Psi^{\,} +\overline\Phi\Psi \Big],
\end{equation}
where $\overline \Phi$ and $\overline \Psi$ are quaternionic conjugates. This real inner product is the foundation of the quantum expectation value in the real Hilbert space $\mathbbm H$QM, and the breakdown of the Ehrenfest theorem in the anti-hermitian approach to  $\mathbbm H$QM  (cf. Section 4.4 of \cite{Adler:1995qqm}) is the physical motivation to  the introduction of the real Hilbert space formalism to $\mathbbm H$QM. The consistency demonstrated in these previous results \cite{Giardino:2016xap,Giardino:2017yke,Giardino:2017nqs,Giardino:2020cee,Giardino:2019xwm,Giardino:2020uab,Giardino:2020ztf} encourage us to apply the real Hilbert space $\mathbbm{H}$QM formalism to quantum systems that do not have satisfactory quaternionic interpretations.

A formal solution to the harmonic oscillator has been sketched in anti-hermitian $\mathbbm{H}$QM 
\cite{Finkelstein:1961tk,Adler:1995qqm}, and a coherent quantization has been obtained in using the regular function approach \cite{Muraleetharan:2014qma,Sabadini:2017qma}. Both of these examples consider the quaternionic Hilbert space, and in this article we use the much simpler approach of the real Hilbert space, and the connection to the $\mathbbm C$QM is accordingly clear and simple. A further example is the biquaternionic harmonic oscillator \cite{Lavoie:2010bcq}.

The article is organized as follows. In Section \ref{U} we revisit the complex result of the infinite square well to obtain the quaternionic
solution. In Secton \ref{T} we repeat the procedure to the finite square well. Section \ref{C} rounds off  the article with our conclusions
and future directions.

\section{\;\sf One-dimensional harmonic oscillator\label{U}}
The quaternionic Schr\"odinger equation for the one-dimensional harmonic oscillator of mass $\mu$ and frequency $\omega$ is simply
\begin{equation}\label{u01}
\hbar\frac{\partial\Psi}{\partial t}i=\left[-\frac{\hbar^2}{2\mu\,}\frac{\partial^2}{\partial x^2}+\frac{1}{2}\mu\omega^2x^2\right]\Psi.
\end{equation}
 The imaginary unit $\,i\,$ multiplies the right hand side of the wave function, and this selection is important in order to define the momentum operator \cite{Giardino:2018lem,Giardino:2018rhs}. Furthermore, although the quaternionic imaginary units are equivalent, only one of them, $\,i,\,$ was elected to  define  the energy and the momentum operators. This common option is important in order to maintain the correspondence between $\mathbbm H$QM and $\mathbbm C$QM. However, a quaternionic theory in which different imaginary units are associated to the energy and momentum operatos is an interesting direction for future research.
The quaternionic wave function $\,\Psi_{nm}\,$ that solves (\ref{u01}) comprises two complex wave functions $\psi_n$, such that
\begin{equation}\label{u02}
\Psi_{nm}=\cos\theta_{nm}\psi_n+\sin\theta_{nm}\overline{\psi}_m\, j,\qquad n,\,m\in\mathbbm{Z}_+,
\end{equation}
where $\,\theta_{mn}\,$ are constants and the  complex wave functions are solutions of the quantum harmonic oscillator ($\mathbbm C$HO). 
 The $\,\theta_{mn}\,$ angle is essential in order to obtain non trivial quaternionic solutions. We will see in a moment   that $\psi_n$ and $\psi_n j$ are orthogonal, despite their identical energies. Consequently, (\ref{u02}) is more constrained than it seems because it expresses the orthogonality requirement for hetero-energetic states.
Thus, let us use the well-known harmonic oscillator solutions of $\mathbbm C$QM
\begin{equation}\label{u03}
\psi_n=\sqrt[4]{\frac{\mu\omega}{\pi\hbar}\,}\frac{1}{\sqrt{2^nn!}\,}H_n(X)e^{-X^2/2}e^{-iE_n t/\hbar},
\qquad{\rm where}\qquad
 E_n=\left(n+\frac{1}{2}\right)\hbar\omega,\qquad X=\sqrt{\frac{\mu\omega}{\hbar}}x,
\end{equation}
and $H_n(X)$ are the Hermite polynomials. The quaternionic harmonic oscillator  solution ($\mathbbm H$HO) given in (\ref{u02}) is not an eigenfunction of the time-independent Schr\"odinger equation, except in the particular case where $n=m$. Therefore, solution (\ref{u02}) describes a coupling between two  complex eigenfunctions of the harmonic oscillator.  The solution must be expressed in a basis for the Hilbert space and a suitable orthogonality condition is needed, a problem that is not solved in the anti-Hermitian case. Applying the definition of the inner product between quaternions (\ref{u003}), we obtain
\begin{equation}\label{u04}
\big\langle\Psi_{nm},\,\Psi_{n'm'}\big\rangle=\cos\theta_{nm}\cos\theta_{nm'}+\sin\theta_{nm}\sin\theta_{n'm}
\end{equation}
where $\langle\psi_n,\,\psi_{n'}\rangle=\delta_{nn'}$ has been used from $\mathbbm{C}$QM. The inner product (\ref{u04}) does not establish the orthogonality between the quaternionic solutions, and an additional constraint is necessary.  Recalling that  $p,\,q\in\mathbbm H$ are parallel ({\em cf.} Section 2.5 of \cite{Ward:1997qcn}) if
\begin{equation}
\mathfrak{Im}[p\bar q]=0,
\end{equation}
we impose the parallelism between the basis elements  as this additional constraint, so that
\begin{equation}
\theta_{nm}=\theta_{n'm'}.
\end{equation}
 Thus, we interpret the angle $\theta_{nm}$ as a parameter that ascribes the degree of interaction between the complex solutions that comprise the quaternionic solution. All the basis elements partake this unique degree of interaction, that we can also understand as polarization of the solution.
Therefore, every element of the basis comprises two polarized wave functions of different energies, and $\theta_{mn}$ is the ``polarization angle'' between these complex components of the quaternionic wave function. Consequently, the condition $\theta_{nm}=\theta_{n'm'}$ sets basis elements of different polarization planes as orthogonal. Accordingly,
\begin{equation}\label{u05}
\big\langle\Psi_{nm},\,\Psi_{n'm'}\big\rangle\,=\,\delta_{nn'}\delta_{mm'}.
\end{equation}
We notice that the pure complex $\,\cos\theta_{nm}\psi_n\,$ and the pure quaternionic $\,\sin\theta_{nm}\psi_m j\,$ components of (\ref{u02}) are mutually orthogonal, in agreement with the interpretation of mechanical polarized waves. 
Afther defining the orthogonality conditions of the wave function, we turn our attention to the expectation values of quaternionic wave functions in a real Hilbert space \cite{Giardino:2018lem,Giardino:2018rhs,Giardino:2019xwm} are obtained from
\begin{equation}\label{u013}
\left\langle\widehat{\mathcal O}\right\rangle= \frac{1}{2}\int dx^3\Bigg[\big(\widehat{\mathcal O}\Psi\big)\overline\Psi^{\,} +\Psi\left(\,\overline{\widehat{\mathcal{O}}\Psi}\,\right) \Bigg],
\end{equation}
and from \cite{Giardino:2019xwm} we know that the expectation values of an arbitrary quaternionic operator $\widehat{\mathcal{O}}$ has the following expression
\begin{equation}\label{q1}
\left\langle\widehat{\mathcal O}_\mathbb{H}\right\rangle=\left\langle\,\widehat{\mathcal O}\right\rangle +\left\langle\big(\widehat{\mathcal O}\,|\,i\big)\right\rangle.
\end{equation}
The contribution of $\big\langle\big(\widehat{\mathcal O}\,|\,i\big)\big\rangle$ is justified physically in order to satisfy the Virial theorem, but this term will not contribute in the case of Hermitian operators. From
a mathematical point of view, this term is the second possibility for defining the scalar product for quaternionic states 
\cite{Harvey:1990sca}, and consequently the expectation value (\ref{q1}) is well defined mathematically.  In the case of Hermitian operators, we get
\begin{equation}\label{u06}
\left\langle\Psi_{nm},\,\widehat{\mathcal{O}}\,\Psi_{nm}\right\rangle\,=\,\cos^2\theta_{nm}\Big\langle\psi_n,\,\widehat{\mathcal{O}}\,\psi_n\Big\rangle\,+\,\sin^2\theta_{nm}\Big\langle\,\overline{\psi}_m,\,\widehat{\mathcal{O}}\,\overline{\psi}_m\Big\rangle,
\end{equation}
 where we used the hermiticity of $\,\widehat{\mathcal O}.\,$ The pure imaginary off diagonal elements cancel out, and the usual complex result is recovered when $\,n=m.\,$ By way of example, the energy expectation value is
\begin{equation}\label{u061}
E_{nm}=\left(n\cos^2\theta_{mn}+m\sin^2\theta_{nm}+\frac{1}{2}\right)\hbar\omega.
\end{equation}
The zero point energy does not suffer any change in the quaternionic formulation, and we can also write the energy as
\begin{equation}\label{u062}
E_{nm}=\left(n+\frac{1}{2}+ (m-n)\sin^2\theta_{nm}\right)\hbar\omega.
\end{equation}
This expression enables us to see the quaternionic part as a correction to the complex part, and the $\theta_{nm}$ angle as the parameter
that regulates the quaternionic influence in the solution. 
The quaternionic solution also admits the algebraic solution of the harmonic oscillator. Using the the operator algebra \cite{Messiah:1999cqm} and the notation $\;(a|b)f=afb\;$ \cite{Giardino:2018lem}, we have
\begin{equation}\label{u07}
\widehat{a}=\frac{1}{\sqrt{2}}\Big[\;X+\left(\widehat{P}\,\big|\,i\right)\,\Big],\qquad
\widehat{a}^\dagger=\frac{1}{\sqrt{2}}\Big[\;X-\left(\widehat{P}\,\big|\,i\right)\,\Big],
\qquad\left[\,\widehat a,\,\widehat a^\dagger\,\right]=1.
\end{equation}
The momentum operator $\widehat{P}$ is such that
\begin{equation}\label{u08}
\widehat P=\frac{1}{\sqrt{\mu\,\hbar\omega\,}}\,\widehat p_x,\qquad\widehat{p}_x=-\hbar(\partial_x|i)\qquad{\rm and}\qquad
\mathcal{H}=\frac{1}{2}\,\hbar\omega\left(\widehat P^2+X^2\right),
\end{equation}
where $\mathcal{H}$ is the Hamiltonian operator of (\ref{u01}). The $\widehat a^\dagger$ is the creation operator, and thus the wave function can be written as
\begin{equation}\label{u09}
\Psi_{nm}=\Big[\,\cos\theta_{mn}A_n e^{-iE_n t/\hbar}\big(\widehat a^\dagger\big)^n\,+\,\sin\theta_{mn}A_m e^{-iE_m t/\hbar}\big(\widehat a^\dagger\big)^m\,j\,\Big]e^{-X^2/2}
\end{equation}
where $A_n$ are normalization constants for $\psi_n$. These results are very simple, and could be easily obtained in the anti-hermitian framework of $\mathbbm{H}$QM. However, the wave funtion (\ref{u02}) is inconsistent in the anti-hermitian
context of $\mathbbm{H}$QM where the orthogonality conditions (\ref{u05}) and the expectation value (\ref{u013}) do not hold and have different definitions. The framework that supports the consistency of the results of this section is the real Hilbert space. The presented results are impossible otherwise and their novelty is totally dependent on it. 
\section{\;\sf   harmonic oscillator in various dimensions\label{T}}
The one-dimensional $\mathbbm H$HO is easily generalized to an arbitrary number $p$ of dimensions, according to
\begin{equation}\label{t01}
\mathcal{H}=\sum_{k=1}^p\mathcal{H}_k,
\end{equation}
where each direction has its own Hamilton operator $\mathcal{H}_k$ that is analogous to (\ref{u08}).
However, the possible solutions for $\mathbbm{H}$QM are much more numerous compared to the $\mathbbm{C}$QM harmonic oscillator. We remember the multi-dimensional harmonic oscillator in $\mathbbm{C}$QM as
\begin{equation}\label{t02}
\psi_n(X)=\prod_{k=1}^p\psi_n^{(k)}(X_k),\qquad\mbox{where}\qquad X=(X_1,\,X_2,\dots X_p)
\end{equation}
and independent oscillations take place along every direction according to the one-dimensional wave function $\,\psi_n^{(k)}\,$. In the quaterninic case, however, there are several possibilities.
 Analogous to (\ref{t02}), we have
\begin{eqnarray}\nonumber
\Psi_{nm}(\bm X)&=&\prod_{k=1}^p \Psi_{nm}^{(k)}(X_k),\\ \label{t03}
&=&\prod_{k=1}^p\left(\cos\theta_{nm}\psi_n^{(k)}+\sin\theta_{nm}\overline{\psi}_m^{(k)}\, j\right),
\end{eqnarray}
where $\,\Psi_k(X_k)\,$ is quaternionic and the total expectation value is the sum of the expectation value at each direction, in complete analogy to the complex case. We observe that the order of the product may change and the energy of the wave function does not change.
A more general possibility for (\ref{t03}) is
\begin{equation}
\Psi_{nm}(\bm X)=\cos\theta_{nm}\prod_{k\in P}\psi_n^{(k)}+\sin\theta_{nm}\prod_{k'\in P'}\overline{\psi}_m^{(k')}\, j\qquad
\mbox{where}\qquad P\cap P'=\{1,\,2,\,\dots p\}.
\end{equation}
This wave function admits much more possibilities than the previous. By way of example, there is a two-dimensional oscillator where the complex and imaginary quaternionic vibrations occur in different directions, and much more possibilities are admitted in higher dimensions.
On the other hand, we may have a third possibility of building a higher dimensional $\mathbbm H$HO using polar coordinates. The time-independent Schr\"odinger equation is
\begin{equation}\label{t05}
\left(-\frac{\hbar^2}{2m}\nabla^2+\frac{1}{2}\mu\,\omega^2 r^2\right)\Phi=E\Phi,
\end{equation}
where $\,\Phi\,$ is a quaternionic wave function. Using spherical coordinates and a radial potential, we have the well known result
\begin{eqnarray}
\label{t06}
\frac{\hbar^2}{2m}\nabla^2_{\hat\theta}\,\mathcal{Y}+\ell\big(\ell+1\big)\mathcal{Y}=0&&\\
\nonumber
\left(-\frac{\hbar^2}{2m}\nabla^2_{\hat r}+\mathcal{V}\right)\mathcal{R}+
\left[\frac{\hbar^2}{2m}\frac{\ell(\ell+1)}{r^2}-E\right]\mathcal{R}=0&&\mbox{where}\qquad
\Phi(r,\,\theta,\,\phi)\,=\,\mathcal{R}(r)\,\mathcal{Y}(\theta,\,\phi).
\end{eqnarray}
The above equations are well known from $\mathbbm{C}$QM but are valid in $\mathbbm{H}$QM as well. 
The real radial solutions of (\ref{t06}) comprise the generalized Laguerre polynomials, $\,L_n^{(\alpha)}(x),\,$ and consequently the quaternionic solutions will be
	\begin{equation}\label{t08}
\mathcal{R}_{uv}(\rho)\,=\,\rho^\ell\, e^{-\rho^2/2}\left[\,\cos\theta_{uv}N_u L_u^{\left(\ell+\frac{1}{2}\right)}\left(\rho^2\right)+\sin\theta_{uv}N_v L_v^{\left(\ell+\frac{1}{2}\right)}\left(\rho^2\right) \,j\,\right]
 \qquad{\rm where}\qquad\rho=\sqrt{\frac{m\omega}{\hbar}}r.
\end{equation}
The normalization constants $N_u$ and $N_v$ of the Laguerre polynomials are known, and also the energy of each  oscillator.
Particularly, the energies are
\begin{equation}\label{t12}
E_{u\ell}=\left(2u+\ell+\frac{3}{2}\right)\hbar\omega\qquad\mbox{where}\qquad u\in\mathbbm{N}.
\end{equation}
In the real Hilbert space, the quaterninic parallelism condition and the orthogonality of the Laguerre polynomials give
\begin{equation}\label{t09}
\big\langle\mathcal{R}_{uv},\,\mathcal{R}_{u'v'}\big\rangle=\delta_{uu'}\delta_{vv'}.
\end{equation}
The radial solution give the energy, and this is absolutely expected considering that the oscillation takes place along the radial direction, and the energy comprises two independent oscillation in the same token as (\ref{u061}). However, we still have a quaternionic solution in the case of $\theta_{uv}=0$. In this specific case, the radial part of the wave function is identical to the complex case and the energy is also identical. On the other hand, the  angular equation of (\ref{t06}) is a combination of spherical harmonics such as 
\begin{equation}\label{t10}
\mathcal{Y}_\ell^{m_1m_2}\big(\theta,\,\phi\big)\,=\,\,\cos\theta_{m_1m_1}Y_\ell^{m_1}\big(\theta,\,\phi\big)\,+\,\sin\theta_{m_1m_1}Y_\ell^{m_2}\big(\theta,\,\phi\big)\,j,
\end{equation}
where $\,Y_\ell^m\,$ is the well known complex spherical harmonic and $\,m_1,\,m_2=\big\{-\ell,\,\dots,\ell\big\}.\,$ The orthogonality condition also take benefit of the parallelism condition to be
\begin{equation}\label{t11}
\Big\langle\,\mathcal{Y}_\ell^{m_1m_2},\,\mathcal{Y}_{\ell'}^{\,m'_1m'_2}\,\Big\rangle\,=\,\delta_{\ell\ell'}\,\delta_{m_1 m_1'}
\,\delta_{m_2 m_2'}.
\end{equation}
As in the complex case, the azimuthal quantum number $\,m\,$ of the spherical harmonic does not contribute to the energy, and this feature is what makes this quaternionic solution possible. The physical properties of the quaternionic spherical harmonic can be further investigated in the scope of the quantum angular momentum and spin.
\section{\;\sf Conclusion\label{C}}
In this article we have provided one of the most important solutions of $\mathbbm{H}$QM in the real Hilbert space: the harmonic oscillator.
The solution of this problem was never obtained in the anti-hermitian version of $\mathbbm{H}$QM, and this fact allows us to suppose that 
the research in real Hilbert space $\mathbbm{H}$QM may have a boost in the future. Almost every application of the harmonic oscillator of $\mathbbm{C}$QM may now be studied using $\mathbbm{H}$QM.  Other fascinating possibilities are the quaternionic version quantum field theory and the supersymmetric quantum mechanics. In both of these the creation-annihilation algebra that has been obtained here will be fundamental.

%
%
%
%

\bibliographystyle{unsrt} 
\bibliography{bib_ohq}

\end{document}